\documentclass[preprint,12pt]{elsarticle}

\usepackage{amsmath}
\usepackage{amsfonts}
\usepackage{amssymb}

\usepackage{color}
\usepackage{bm}
\usepackage{subcaption}
\usepackage{multirow}

\journal{Ocean Engineering}

\begin{document}

\begin{frontmatter}



\title{Artificial neural networks ensemble methodology to predict significant wave height}


\author[inst1]{Felipe Crivellaro Minuzzi}
\author[inst2,inst3,inst4]{Leandro Farina}

\affiliation[inst1]{organization={Departament of Mathematics, Federal University of Santa Maria (UFSM)},
            addressline={Av. Roraima 1000}, 
            city={Santa Maria},
            postcode={97105-900}, 
            state={RS},
            country={Brazil}}

\affiliation[inst2]{organization={Institute of Mathematics and Statistics, Federal University of Rio Grande do Sul (UFRGS)},
            addressline={Av. Bento Goncalves 9500},
            city={Porto Alegre},
            postcode={91509-900}, 
            state={RS},
            country={Brazil}} 

\affiliation[inst3]{organization={Center for Coastal and Oceanic Geology Studies (CECO), Federal University of Rio Grande do Sul (UFRGS)},
            addressline={Av. Bento Goncalves 9500},
            city={Porto Alegre},
            postcode={Bulding 43.125}, 
            state={RS},
            country={Brazil}}
            
\affiliation[inst4]{organization={Postgraduate Program in Geosciences, Federal University of Rio Grande do Sul (UFRGS)},
           addressline={Av. Bento Goncalves 9500},
            city={Porto Alegre},
            postcode={91501-970}, 
            state={RS},
            country={Brazil}}

\ead{felipe.minuzzi@ufsm.br}
\ead{farina@alum.mit.edu}

\begin{abstract}
The forecast of wave variables are important for several applications that depend on a better description of the ocean state. Due to the chaotic behaviour of the differential equations which model this problem, a well know strategy to overcome the difficulties is basically to run several simulations, by for instance, varying the initial condition, and averaging the result of each of these, creating an ensemble. Moreover, in the last few years, considering the amount of available data and the computational power increase, machine learning algorithms have been applied as surrogate to traditional numerical models, yielding comparative or better results. In this work, we present a methodology to create an ensemble of different artificial neural networks architectures, namely, MLP, RNN, LSTM, CNN and a hybrid CNN-LSTM, which aims to predict significant wave height on six different locations in the Brazilian coast. The networks are trained using NOAA's numerical reforecast data and target the residual between observational data and the numerical model output. A new strategy to create the training and target datasets is demonstrated. Results show that our framework is capable of producing high efficient forecast, with an average accuracy of $80\%$, that can achieve up to $88\%$ in the best case scenario, which means $5\%$ reduction in error metrics if compared to NOAA's numerical model, and a increasingly reduction of computational cost.
\end{abstract}


\begin{keyword}
Ocean modelling \sep Machine learning \sep Artificial neural networks \sep Significant wave height \sep  Forecast
\end{keyword}

\end{frontmatter}

\section{Introduction}\label{intro}


Numerical simulations of both weather and ocean parameters rely on the evolution of nonlinear dynamical systems that have a high sensitivity on initial conditions. Considering that errors in the observations and analysis are present, and therefore in the initial conditions, the concept of a unique deterministic solution of the governing equations becomes fragile \cite{komen1996dynamics, farina2002ensemble}. To circumvent this drawback, one can use an ensemble of simulations with different initial conditions, to represent the uncertainty of the data, and generates different solutions in which its average can provide a better understanding of the medium range behaviour of the system.

Albeit mathematical-physical models can be solved using traditional numerical solvers \cite{wavewatch, booij1997swan}, the amount of available quality data prompt the use of machine learning algorithms as an low-cost alternative, achieving better performance in a computational time that is incredibly reduced.  
In this sense, artificial neural networks (ANNs) are one of the most promising tools for numerical simulations and act as an important alternative to problems with random patters such as those found in ocean modelling \cite{juan2022review, minuzzi2023deep}.

Artificial neural networks are a class of supervised learning algorithms that produce data-driven results, i.e., the machine has the ability to learn non-linear patterns from a huge amount of data. It is designed to model the way in which the brain performs a particular task \cite{haykin2009neural} and, from a statistical and mathematical standpoint, it is a multiple nonlinear regression method mapping inputs with outputs.

Since a ensemble prediction system (EPS) is a well established framework to predict ocean wave physical variables \cite{farina2002ensemble, bunney2015ensemble, behrens2015development, saetra2004potential}, recently, several works employed the ensemble methodology with artificial neural networks \cite{krasnopolsky2012neural, campos2022mid}. O'Donncha \textit{et al.} \cite{o2019ensemble} applied a methodology to create a machine learning algorithm that combined forecasts from multiple, independent models into an ensemble prediction of the true state, while Gr{\"o}nquist \textit{et. al} \cite{gronquist2021deep} proposed a model that uses a subset of the original weather trajectories together with a post-processing step using neural networks and achieved an improvement in the ensemble forecast of about $14\%$, with large metrics for extreme weather events. Gao \textit{et al.} \cite{gao2023significant} proposed a model to predict significant wave height that consists in an ensemble of different random vector functional link (RVFL) deep learning algorithms and the traditional ARIMA method. Xu \textit{et al.} \cite{xu2022novel} developed wave-induced forces predictions using an ensemble of four state-of-the-art surrogate models. The final results are obtained with a weighted average defined by an artificial neural network and it was shown that the proposed methodology capable of providing robust and accurate approximation for different force components.

Campos \textit{et al.} \cite{campos2020improving} developed a post-processing algorithm to improve ensemble averaging, as a replacement to the typical arithmetic ensemble mean, using neural networks trained with altimeter data, that predicted the residual of significant wave height ($H_s$) and wind speed ($U_{10}$), i.e, the difference from the ensemble average and the observations. In a previous paper, Campos \textit{et al.} had already used the methodology of training neural networks to predict the residual signal applied to a global wave ensemble (as a hybrid approach), trained using buoy data \cite{campos2019nonlinear}, although not using ensemble framework. The residual of $H_s$ was also predicted in the work of Marangoni \cite{marangoni2023predicting} and collaborators, where the long short-term memory (LSTM) architecture of neural network was used as post-processing model to improve outputs from the numerical model. A comprehensive review on artificial neural networks on ocean engineering can be found in \cite{juan2022review}. Furthermore, several machine learning techniques have been applied to oceanography predictions, such as transformer neural networks \cite{pokhrel2022transformer}, support vector machines~\cite{browne2007near, ccelik2022improving}, Bayesian optimization \cite{cornejo2018bayesian, adnan2023short}, genetic programming~\cite{nitsure2012wave,gaur2008real} and wavelets~\cite{oh2018real,prahlada2015forecasting,altunkaynak2023hourly}, among others \cite{altunkaynak2015prediction, ikram2023improving, mostafa2023modeling, sadeghifar2022wave, sadeghifar2017coastal}, with results that outperform traditional models, in several specific cases. 

The prediction of a variable residual instead of its actual value has an advantage when using neural networks within ensemble predictions. As the final result will be added to the ensemble mean (EM), in the training process, the parameters of the activation function will only update the EM deviates from the target value \cite{campos2020improving, campos2019nonlinear}.

Benefiting from both the advantages of using ensemble and artificial neural networks, we aim to provide in this work a new methodology to forecast significant wave height $H_s$ on six different locations in Brazilian coast. We build five different architectures of artificial neural networks in which each predicts the residual between the observed value and $H_s$ output from a numerical model. The final result is calculated by averaging $H_s$ from the different architectures, which is reconstructed by adding a forecast $H_s$ with the neural network residual. To obtain such predictions, a novel methodology to build a specific training and testing datasets is presented. The data is obtained both from real observed values and numerical simulations outputs.

This work is established as follow:  in Section \ref{ens}, we describe the five different architectures of neural networks used. In Section \ref{metho}, we present the framework to construct the training and target datasets that will be used for simulation, as well as the particularities of these. In Section \ref{data}, the data and area of study is explained, followed by Section \ref{results}, with the results and discussion. Finally, the conclusions are presented in Section \ref{conclusion}.

\section{Ensemble of neural networks} \label{ens}


Artificial neural networks (ANNs) are a kind of machine software  that is designed to model the way in which the brain performs a particular task, and is able to learn and generalize huge sets of data \cite{haykin2009neural}. From a mathematical standpoint, they can be considered as multiple nonlinear regression methods able to capture hidden complex nonlinear relationships between input and output variables \cite{peres2015significant}.

In its simplest form, know as the perceptron, the structure of an ANN is based on a unit, or neuron ($y_k$), which receives a linear combination of weighted input and bias, i.e \cite{haykin2009neural, makarynskyy2004improving},
\begin{eqnarray}
y_k & = & \phi \left(\sum_{j=1}^m \omega_{kj}x_j + b_k \right) \label{percep}
\end{eqnarray} where $\omega_{kj}x_j$ for each $j$ consists of the multiplication of the synaptic weight $\omega_{kj}x_j$ and the data $x_j$ and $b_k$ indicates the bias, which has the effect of increasing or lowering the net input of the activation function $\phi$. As we aim to make our network accountable for non-linear dependencies, the activation functions need to be also non-linear, such as the log sigmoid or the hyperbolic tangent sigmoid functions. Nevertheless, this choice is user-defined and may depended on the application.

The determination of the weights and biases in the network is executed in the learning phase of the algorithm, and they are adjusted iteratively based on the data given as input-output that are seen by the network. This process aims at minimize the loss (or performance) function which can be, for instance, the squared error between the output of the network and the real output value, in order to make the network perform in an expected way. The widely used framework in the learning algorithm of an ANN is the gradient descent backpropagation, which updates weights and biases in the direction of the negative gradient of the loss function. 

Several artificial neural network (ANN) architectures, based on layers of neurons, are possible. Different approaches to how information circulates throughout the network are also possible. In this work, we construct five different architectures of neural networks and average the results of each of these, to construct an ensemble of neural networks. In what follows, we briefly describe the characteristics of each architecture. We invite the reader to the references \cite{haykin2009neural, goodfellow2016deep, krasnopolsky2013application} to a full description of the artificial neural networks used in this work.

\subsection{Multilayer perceptron}

The structure known as multilayer perceptron (MLP) is a construction that unites several hidden layers to map the input and output layers. Each of these has a number of neurons that simulates an output by means of Eq. \eqref{percep}. This network exhibits a high degree of connectivity \cite{haykin2009neural}, determined by the weights of the network.

Mathematically, with the same notation used above in Eq. \eqref{percep}, if a MLP networks has $H$ hidden layers, $n$ inputs and $m$ outputs, then an output $y_k$, where $k = 1, \ldots, m$, is given by \cite{krasnopolsky2013application}
\begin{eqnarray}
    y_k & = & b_{k} + \sum_{j = 1}^H \omega_{kj} \cdot \phi \left( \sum_{i = 1}^n \omega_{ji}x_i + b_k \right) 
\end{eqnarray}

Although a simple structure, MLPs has already demonstrated high accuracy for modelling ocean wave variables, such as significant wave height \cite{londhe2006one}. 

\subsection{Recurrent neural networks}

Recurrent neural networks (RNNs) take advantage over MLP by sharing parameters across different parts of a model \cite{goodfellow2016deep}. RNNs can be derived from nonlinear first-order non-homogeneous ordinary differential equations and the idea is to store some information about the past time evolution of the system in a hidden state vector, which means that a neuron's output can be feedbacked as an input to all neurons of the net. This attribute of RNNs allows a memory of previous inputs to persist in the network’s internal state, and thereby influence the network output \cite{graves2008supervised}. The forward propagation equations for a RNN, with the specification of the initial state ${\bm h}^0$, from time $t$ to time $\tau$, can be given by \cite{goodfellow2016deep}
\begin{eqnarray}
    {\bm a}^t & = & {\bm b} + {\bm W} {\bm h}^{(t-1)} + {\bm U} {\bm x}^t, \\
    {\bm h}^t & = & \tanh{({\bm a}^t)}, \\
    {\bm o}^t & = & {\bm c} + {\bm V} {\bm h}^t,
\end{eqnarray} where ${\bm b}$ and ${\bm c}$ are the bias vectors and ${\bm U}$, ${\bm V}$ and ${\bm W}$ the weight matrices for input-to-hidden, hidden-to-output and hidden-to-hidden, respectively.

Nevertheless, the use of RNNs in its standard configuration to account for contextual information is still limited, due to the vanishing gradient problem \cite{haykin2009neural}, i.e., as the information circles around the recurrent network in time, the influence of a input on the hidden layer, and consequently on the output, either decays or blows up exponentially. One attempt to solve this problem is with the long short-term memory (LSTM) architecture, presented for the first time by Hochreiter and Schmidhuber \cite{hochreiter1997long}, exploited in the next section.

\subsection{Long short-term memory}

Long short-term memory (LSTM) architecture incorporates non-linear data-dependent controls into the RNN cell in order to solve the problem of the vanishing gradient\cite{sherstinsky2020fundamentals}. The main difference between LSTMs and RNNs is that the summation in the hidden layer is replaced by a memory block, which has four neural networks connected and interacting together. This structure allows LSTMs to learn and remember information for a long-time period, which is its default behaviour \cite{minuzzi2023deep}.

The architecture of LSTMs is build as follow: inside the memory block, there is one or more central cells that are self-looped into three multiplicative units called input, output and the forget gates. This difference of having more units controls the flow of information \cite{goodfellow2016deep}, where the multiplicative input protects the memory block from receiving perturbation from irrelevant inputs, while the output gates protect other units from irrelevant information of the current block \cite{hochreiter1997long}. The units work as gates to avoid weight conflicts, i.e., the input gate decides when to keep or exclude information within the block, while the output gate decides when to access the block and prevent other blocks from being perturbed by itself. If $n$, $m$ and $k$ correspond to the number of inputs, outputs and cells in the hidden layer, respectively, the equations that formulate a LSTM network are given by \cite{hochreiter1997long}
\begin{eqnarray}
b_{\sigma}^t & = & \phi \left(  \sum_{i = 1}^n \omega_{i \sigma}x_i^t + \sum_{h = 1}^m \omega_{h \sigma}b_h^{t-1} + \sum_{c = 1}^k \omega_{c \sigma}s_c^{t-1}  \right), \\
b_\tau^t & = & \phi \left(  \sum_{i = 1}^n \omega_{i \tau}x_i^t + \sum_{h = 1}^m \omega_{h \tau}b_h^{t-1} + \sum_{c = 1}^k \omega_{c \tau}s_c^{t-1}  \right), \\
b_\gamma^t & = & \phi \left(  \sum_{i = 1}^n \omega_{i \gamma}x_i^t + \sum_{h = 1}^m \omega_{h \gamma}b_h^{t-1} + \sum_{c = 1}^k \omega_{c \gamma}s_c^{t}  \right), \\
s_c^t & = & b_\tau^t s_c^{t-1} + b_{\sigma}^t g \left(   \sum_{i = 1}^n \omega_{i c}x_i^t + \sum_{h = 1}^m \omega_{h c}b_h^{t-1}  \right), \\
b_c^t & = & b_\gamma^t h(s_c^t),
\end{eqnarray}where $b_{\sigma}$ represents the input gate, $b_\tau$ the forget gate, $b_\gamma$ the output gate, $x$ the signal, $\omega$ the weights that will connect two units and $s_c^t$ is the activation of the cell $c$ at time $t$ unit. In the formulae above, $\omega_{c \sigma}$, $\omega_{c \tau}$ and $\omega_{c \gamma}$ indicate the weights from the cell to the input, forget and output gates, respectively.

\subsection{Convolutional neural network}

Convolutional neural networks (CNNs) are a specialized kind of neural network for processing data that has a known grid-like topology \cite{goodfellow2016deep}, which make them suitable for a bi-dimensional forecast of sea state, for instance. The name indicates that the network employs a mathematical operation called convolution, i.e., a specialized kind of linear operation, given by
\begin{eqnarray}
(x \ast w)(t) & = & \int x(a)w(t-a) da
\end{eqnarray} where, in a machine learning problem, $x$ is the input of data and $w$ is referred as the kernel. The output of this operation is usually called feature map. As to implement this methodology in the computer, time must be discretized. Therefore, assuming $x$ and $w$ are only functions of $t$ a discrete convolution is defined as
\begin{eqnarray}
(x \ast w)(t) & = & \sum_{- \infty}^{\infty}x(a)w(t-a).
\end{eqnarray} Convolution operations improve learning processes due to their sparse interactions, parameter sharing and equivariant representations \cite{goodfellow2016deep}.

In this type of architecture, at least one layer uses convolution in place of general matrix multiplication, which consists of three stages \cite{goodfellow2016deep}: in the first stage, the layer carries out several convolutions operations in parallel to produce a set of linear activations. Then, in the second stage, each of these goes through a nonlinear activation function and in the third stage, a pooling function is used to replace the output of the net at a certain location with a statistic value of the nearby outputs. The pooling operation helps to make the representation invariant to small modifications of the input. A convolutional filter is essentially a weighting vector/matrix/cube that uses a sliding-window approach \cite{reichstein2019deep}.

\subsection{Hybrid CNN with LSTM}

While CNN has the power to treat grid-like structures and learn the correlations between neighbor points, LSTM can help to learn long-term patterns in the data. Therefore, both these networks can be used together to forecasting problems. A convolutional LSTM network has been recently been built for precipitation nowcasting \cite{xingjian2015convolutional}, a model that has been shown to outperform traditional optical flow based methods. Thus, we propose a hybrid CNN-LSTM network, where, in the first phase, a CNN is applied for features extraction on input data, and LSTM to interpret this features and develops a sequence prediction.

\section{The ensemble methodology} \label{metho}

As the use of neural networks to predict a residual has already been discussed and applied with satisfactory results \cite{campos2020improving,campos2019nonlinear}, we propose a different methodology to construct the datasets that will be used to train each of the neural networks mentioned in the previous section. The target, i.e., the variable that will be predicted is the residual of the significant wave height $H_s$, calculated as the difference between the real observed value and the forecast output of a numerical model. We consider the net residual, and not the absolute values, to account for negative values.

As the output of the neural networks consist of residuals, we reconstruct $H_s$ by adding these residuals to a numerical prediction of $H_s$ for the respective forecast horizon. We opted for this framework because allows us to generate an operational forecast that can be used on a daily basis. Afterwards, an average of the five $H_s$ results is calculated which yields the final result of the algorithm's prediction.

\subsection{Creating the training datasets} \label{training}

\begin{figure}[!ht]
\begin{center}
    \includegraphics[width=\textwidth]{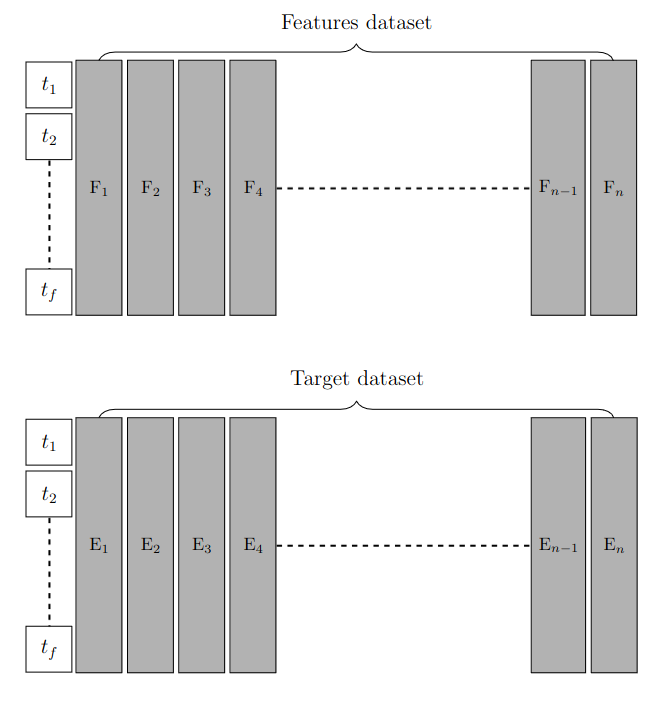}
\end{center}
\caption{Schematic of the methodology for training applied in this work. Here, $t_1, \ldots, t_f$ represents the time series. In the target dataset, E$_i$ represents the different between the numerical model forecasts from columns F$_i$ and buoy data. The F$_i$ columns are numerical model forecast on a specific lead time (column $i = 1$, lead time is 3hrs, $i = 2$, lead time is 6hrs, and so on until the $n-$th column).}
\label{fig:1}
\end{figure}

Figure \ref{fig:1} shows a schematic of the training methodology developed. In this framework, we build a feature dataset, in which each column have a numerical time series forecast for a specific lead time. First column contains data from 3-hour lead forecasts of $H_s$, column two, 6-hour lead forecasts, and so on until the $n-$th column. Each row of this dataset represents a date and time, and its length define the size of the training phase. For the target dataset, each element $(i,j)$ will be the residual between the numerical model in the features dataset at position $(i,j)$ and the real measured data obtained from the buoy at that date and time. In this sense, the predictions of the neural network will be of one row and $n$ columns, in the time position immediately following the last row of the target dataset. Therefore, the network will have predicted the residual, and considering each of the lead times of the columns, as it was constructed in the features dataset.

In the training phase of every neural network, a cross-validation scheme was implemented, where $80\%$ of the data is selected for the training and $20\%$ for validation. This strategy is an excellent framework to avoid overfitting of a model, i.e., a model that yields a good accuracy to the validation set (seen data) and a bad result to unseen data. Since we are training with time series data, the order of events is important, which can be a problem when using cross-validation. To circumvent this issue, we perform a cross-validation on a rolling basis, where the training dataset is divided into smaller batches of data, and the cross-validation is applied to these batches. We train in a subset of data and then forecast the later data points of the batch to check accuracy. The same forecasted data points are then included as part of the next batch of training. This strategy also avoid excess in the memory usage of the training phase. To define the batch size, several simulations were performed, and a optimal value of twelve data points was obtained.

The Python library TensorFlow \cite{tensorflow2015-whitepaper} is an end-to-end open source platform for machine learning and its Keras API \cite{chollet2015keras} are used in this work to implement the neural networks. The model is compiled using the mean absolute error as loss function which is optimized by the Adam algorithm. Adam optimization is a stochastic gradient descent method that is based on adaptive estimation of first-order and second-order moments. The networks are build with six hidden layers (the hybrid CNN-LSTM has six hidden layers for each of the architectures) and the hyperbolic tangent is used as activation function, to account for negative values of the residual. A similar structure of simulation had already been used with satisfying results \cite{minuzzi2023deep}. In order to allow reproducibility, the code used in the simulations and also the data can be found in \cite{minuzzigit}.

\subsection{Error metrics}

In this work, to evaluate the performance of our proposed model in comparison with the true observational data, we use four metrics to analyse the accuracy of our results. The well known mean absolute error (MAE), which is given by
\begin{eqnarray}
MAE & = & \frac{1}{n} \sum_{i = 1}^n \left| \tilde{y}_i - y_i \right|,
\label{eq1}
\end{eqnarray} where the tilde means reference value while non-tilde means predicted value ($n$ is the number of observations). The root mean squared error (RMSE), to analyse how disperse the error from the real value our ensemble methodology is, given by
\begin{eqnarray}
RMSE & = & \sqrt{ \dfrac{1}{n} \sum_{i = 1}^n ( \tilde{y}_i - y_i)^2}.
\label{eq2}
\end{eqnarray} We also show the comparison through relative error (RE), a percentage measure of the difference from the actual value, given point-by-point in the domain of forecast,
\begin{eqnarray}
RE & = & 100 \times \frac{\left| \tilde{y}_i - y_i \right|}{\left| \tilde{y}_i \right|},
\label{eq3}
\end{eqnarray}
and the mean absolute relative error (MAPE), which is an average of the relative error,
\begin{eqnarray}
MAPE & = & 100 \times \frac{1}{n} \sum_{i = 1}^n \frac{\left| \tilde{y}_i - y_i \right|}{\left| \tilde{y}_i \right|}.
\label{eq4}
\end{eqnarray}

\section{Data and area of study} \label{data}

The objective of our work is to forecast significant wave height $H_s$. The framework described in the previous section is used to predict the residue between numerical and observational $H_s$ data, which later is reconstructed by adding these residuals to a numerical prediction of $H_s$. We use NOAA Wave Ensemble Reforecast data \cite{reforecast} as input, which is a 20-year global wave reforecast generated by the WAVEWATCH III model \cite{wavewatch}, forced by NOAA's Global Ensemble Forecast System (GEFSv12) \cite{gefs}. The wave ensemble was run with one cycle per day, spatial resolution of $0.25^o \times 0.25^o$ and temporal resolution of three hours. More details can be found in the documentation \cite{reforecast}. The forecast range is sixteen days, which is also the same range in which we perform the forecast using the neural networks in this work. We use only the deterministic member of the NOAA's reforecast data. 

\begin{table}
\scriptsize
    \begin{tabular}{l|c|c|c|c|c|c}
         & Longitude & Latitude & Period of prediction & Depth (m) & WMO & City/State location \\ \hline
        Buoy 1 & 49° 86' W & 31° 33' S  & 20/02/2019 -- 08/03/2019 &  200 &  31053 & Rio Grande/RS \\
        Buoy 2 & 47° 15' W & 27° 24' S  & 30/10/2018 -- 15/11/2018 & 200  & 31231 & Itajaí/SC \\
        Buoy 3 & 42° 44' W & 25° 30' S  & 28/04/2018 -- 14/05/2019 & 2164 & 31374 & Santos/SP \\
        Buoy 4 & 39° 41' W & 19° 55' S  & 06/07/2017 -- 22/07/2017 & 200 & 31380 & Vitória/ES \\
        Buoy 5 & 34° 33' W & 8° 09' S & 31/10/2015 -- 16/11/2015 & 200 & 31229 & Recife/PE \\
        Buoy 6 & 38° 25' W & 3° 12' S & 08/04/2018 -- 24/04/2018 & 200  & 31229 & Fortaleza/RN \\
        \hline
    \end{tabular}
    \caption{Geo-spatial latitude and longitude location of the six buoys used in this work, period of prediction, water depth, WMO identification number and city of location in Brazil.}
    \label{tab:1}
\end{table}

\begin{figure}
    \centering
    \includegraphics[width=\textwidth]{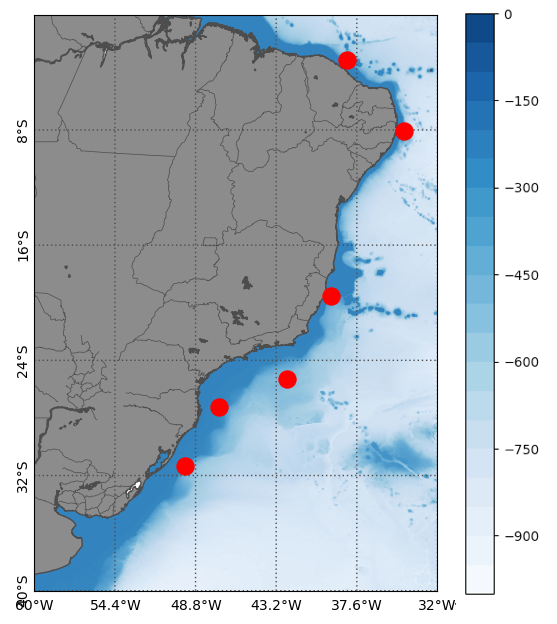}
    \caption{Region of analysis in the Brazilian coast with bathymetry. Red circles indicates the location of the six buoys studied in this work. The blue color scale represents the bathymetry (in meters).}
    \label{fig:2}
\end{figure}

Data from six buoys are also used for this study. All of them are located in the Brazilian coast, ranging from longitude 49° 86'W to 38° 25'W and latitudes 31° 33'S to 3° 12'S. These buoys belong to the National Program of Buoys (PNBOIA) of the Brazilian Navy, which aims to collect oceanographic and meteorological data of the Atlantic Ocean \cite{pereira2017wave, pnboia}, and a detailed description of the data quality can be found in \cite{pnboiaquality}. We interpolated data for missing points in these datasets. The training (and consequently the prediction) period is also determined to be the one with the least missing points. Figure~\ref{fig:2} shows the region of analysis, with red circles indicating the location of the buoys, while Tab. \ref{tab:1} presents the longitudes and latitudes of the six buoys. One limitation of our work can be inferred from Table \ref{tab:1}, which shows the depth, in meters, of the buoys locations. Depending on the position, these can be considered coastal, which is not the goal of the NOAA Wave Ensemble, designed for deep waters. We present data statistics for both the real observations and NOAA ensemble numerical model in Tab. \ref{tab:statistics}.

The prediction period varies for each buoy, since some of the buoys used in this work are on maintenance and do not have real time data. We gathered this information for each buoy in Table~\ref{tab:1}. The training period is from 2013 until each buoy's prediction starting date. The results are shown for every three hours, the same temporal resolution of the reforecast data.
\begin{table}
\scriptsize
    \centering
    \begin{tabular}{l|cccccccccc}
    \hline
    \multirow{ 2}{*}{Data statistics} &
    \multicolumn{2}{c}{Mean}  &
    \multicolumn{2}{c}{Standard Deviation} &
    \multicolumn{2}{c}{Q1} & 
    \multicolumn{2}{c}{Q2} &
    \multicolumn{2}{c}{Q3}  \\
     &  Buoy data & NOAA &  Buoy data & NOAA &  Buoy data  & NOAA &  Buoy data & NOAA &  Buoy data & NOAA  \\
     \hline
     Rio Grande & 1.84 & 1.77 & 0.79 & 0.50 & 1.21 & 1.40 & 1.64 & 1.73 & 2.45 & 2.08 \\
     Itajaí & 1.05 & 1.82 & 0.49 & 0.27 & 1.67 & 1.63 & 1.87 & 1.76 & 2.44 & 2.03 \\
     Santos & 1.67 & 1.85 & 0.43 & 0.60 & 1.25 & 1.31 & 1.63 & 1.69 & 2.07 & 2.21 \\
     Vitória & 1.79 & 1.98 & 0.37 & 0.44 & 1.54 & 1.61 & 1.77 & 2.00 & 1.99 & 2.36 \\
     Recife & 1.48 & 1.79 & 0.27 & 0.30 & 1.28 & 1.57 & 1.47 & 1.86 & 1.62 & 1.98 \\
     Fortaleza & 1.45 & 1.56 & 0.27 & 0.28 & 1.28 & 1.35 & 1.48 & 1.56 & 1.63 & 1.73 \\
    \hline
    \end{tabular}
    \caption{Real buoy observations and NOAA ensemble forecast data statistics. All values are given in meters. Q1, Q2 and Q3 stand for first quartile, second quartile and third quartile, respectively.}
    \label{tab:statistics}
\end{table}

\section{Results and discussion} \label{results}

In this section, we present the results of the prediction carried out with the ensemble of artificial neural networks that was described above (referred as NN ensemble in what follows). The residual that is the target of each simulation is added to a numerical forecast from NOAA Wave Ensemble Reforecast. All the metrics are calculated against buoy data observations. 

\begin{figure}[!ht]
     \centering
     \begin{subfigure}{\textwidth}
         \centering
         \includegraphics[width=\textwidth]{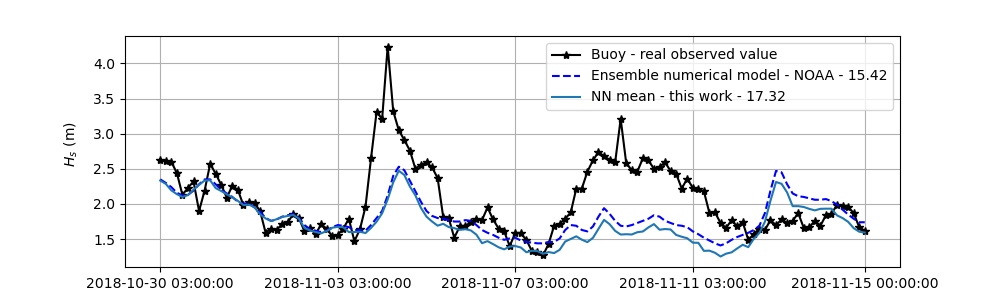}
     \end{subfigure}
     \hfill
     \begin{subfigure}{\textwidth}
         \centering
         \includegraphics[width=\textwidth]{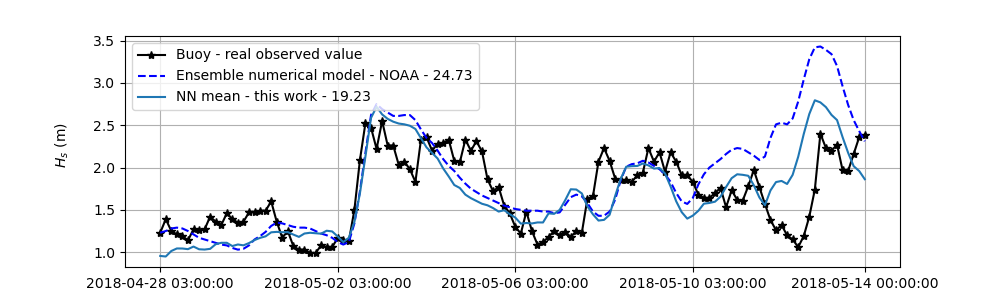}
     \end{subfigure}
     \hfill
     \begin{subfigure}{\textwidth}
         \centering
         \includegraphics[width=\textwidth]{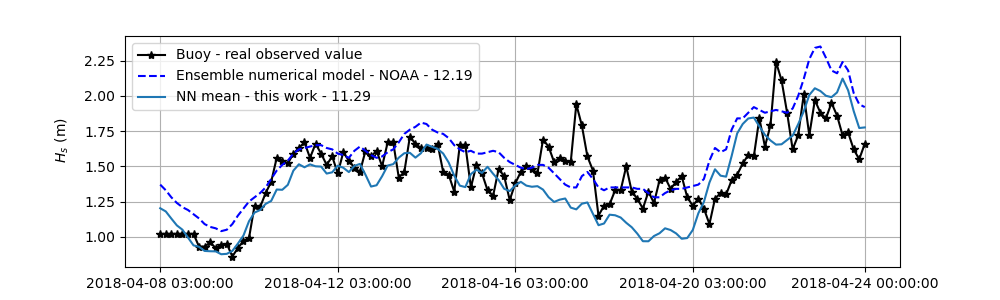}
     \end{subfigure}
        \caption{Comparison between real observed data, NOAA reforecast numerical simulation and this work ensemble methodology. Numbers in the legend refers to MAPE metric. Upper: Itajaí, Middle: Santos, Bottom: Fortaleza.}
        \label{fig:3}
\end{figure}

\begin{figure}[!ht]
     \centering
     \begin{subfigure}{\textwidth}
         \centering
         \includegraphics[width=\textwidth]{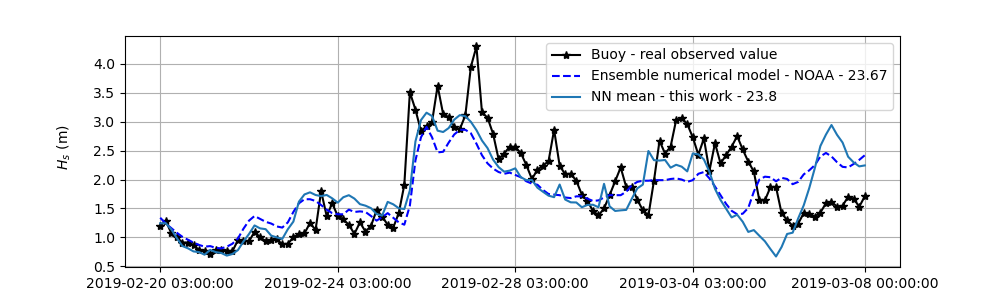}
     \end{subfigure}
     \hfill
     \begin{subfigure}{\textwidth}
         \centering
         \includegraphics[width=\textwidth]{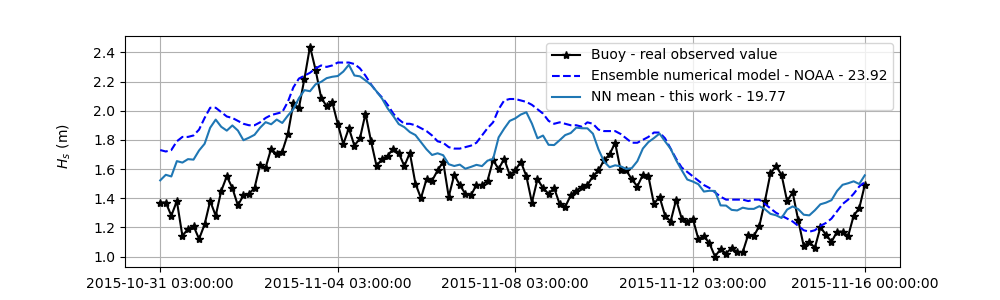}
     \end{subfigure}
     \hfill
     \begin{subfigure}{\textwidth}
         \centering
         \includegraphics[width=\textwidth]{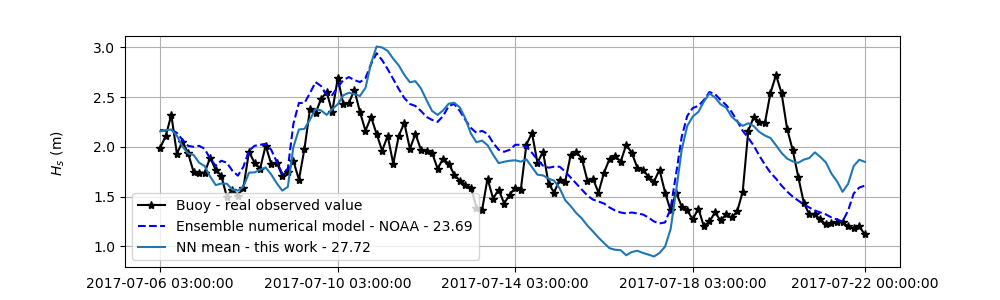}
     \end{subfigure}
        \caption{Comparison between real observed data, NOAA numerical simulation and this work ensemble methodology. Numbers in the legend refers to MAPE metric. Upper: Rio Grande, Middle: Recife, Bottom: Vitória.}
        \label{fig:4}
\end{figure}

Figures \ref{fig:3} and \ref{fig:4} illustrate the comparison results between observed data, NOAA reforecast numerical model and this work ensemble of neural networks. As we can see, there is no quantitative improvement in the MAPE metric if we compare the numerical model and the neural networks ensemble. Buoys locations at Santos (Fig \ref{fig:3}, middle), Recife and Vitória (Fig. \ref{fig:4}, middle and bottom) show the greatest discrepancy; the first and the second with a better accuracy for the NN ensemble while the third with a better accuracy for the NOAA numerical model. 

Figure \ref{fig:4} highlights how the NN ensemble fails to predict $H_s$ peaks in buoy location at Itajaí. The reason might be because we are training the models and reconstructing $H_s$ with data from NOAA numerical simulation and since global wave models are known to not represent extreme events very well, the pattern is also learned by the neural networks. The poor representation of peaks is also seen in other buoy locations, and this could be addressed as one of the drawbacks of the proposed methodology. It is well known that, in the numerical forecast of ocean waves, peaks, storms and extremes events are difficult to predict. The training of the neural networks are based on the NOAA numerical results, which can explain the limitation. Also, data imbalance, since peaks and storms represent a small portion of the training dataset, as well as the use of a numerical global model, are others problems that prevents from getting better results.

One can also see that the ensemble in fact learn and predict a residual that is variable according with the initial error, as the graphs show, and although in the beginning of the prediction both numerical and NN have the same behaviour, later on the prediction period the lines get apart from each other, specially where the numerical model is known to lose accuracy, the ensemble of neural networks maintains it. The results show also the same pattern of balance in the error if one looks at the metrics MAE and RMSE, as can be seen in Tab. \ref{tab:3}. However, the cost of simulation for the ensemble is vastly reduced compared to the numerical simulation, which can be seen as an advantage. For each neural network architecture, our algorithm took approximately 32 minutes for training and the prediction of a single time value took $2.62 \times 10^{-6}$ seconds. Thus, the sixteen days predictions period (128 steps) took $3.35 \times 10^{-4}$ seconds. The simulations were performed in a machine with Intel Xeon processor with $20$ cores, $128$ Gb of RAM memory, with a GeForce RTX 2080 Ti graphics card. We parallelize all the training and prediction step, so the results for each architecture are given in the same time.

\begin{table}
\scriptsize
    \centering
    \begin{tabular}{l|cccccccccccc}
    \hline
    \multirow{ 2}{*}{Error metrics} &
    \multicolumn{2}{c}{Itajaí}  &
    \multicolumn{2}{c}{Santos} &
    \multicolumn{2}{c}{Fortaleza} & 
    \multicolumn{2}{c}{Rio Grande} &
    \multicolumn{2}{c}{Recife} &
    \multicolumn{2}{c}{Vitória} \\
     &  NN & NOAA &  NN & NOAA &  NN  & NOAA &  NN & NOAA &  NN & NOAA &  NN & NOAA \\
     \hline
     MAPE ($\%$) & 17.32 & 15.42 & 19.23 & 24.73 & 11.29 & 12.19 & 23.8 & 23.67 & 19.77 & 23.92 & 27.72 & 23.69 \\
     MAE ($m$)   & 0.40   & 0.40  & 0.30  & 0.39  & 0.17  & 0.17  & 0.44 & 0.44  & 0.28  & 0.34  & 0.45  & 0.39  \\
     RMSE ($m$)  & 0.56  & 0.56  & 0.38  & 0.55  & 0.21  & 0.22  & 0.58 & 0.57  & 0.31  & 0.38  & 0.56  & 0.50  \\
    \hline
    \end{tabular}
    \caption{Error metrics for each buoy locations. Comparison between ensemble neural network and NOAA numerical model against real observational data.}
    \label{tab:3}
\end{table}

\begin{figure}[!ht]
     \centering
     \begin{subfigure}{\textwidth}
         \centering
         \includegraphics[width=\textwidth]{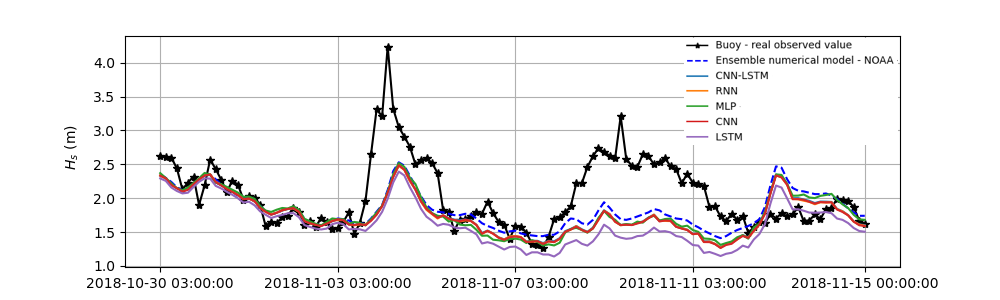}
     \end{subfigure}
     \hfill
     \begin{subfigure}{\textwidth}
         \centering
         \includegraphics[width=\textwidth]{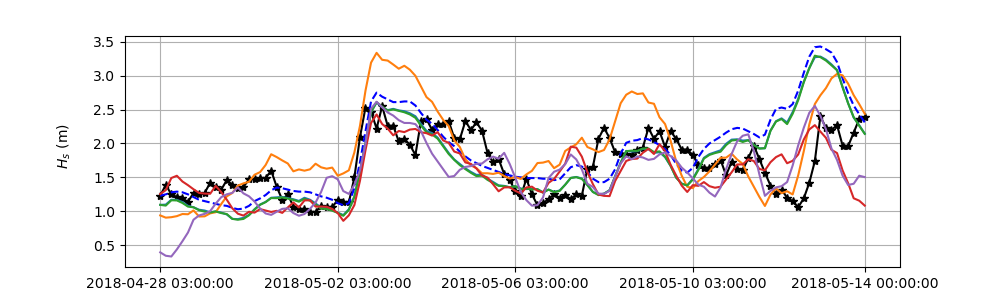}
     \end{subfigure}
     \hfill
     \begin{subfigure}{\textwidth}
         \centering
         \includegraphics[width=\textwidth]{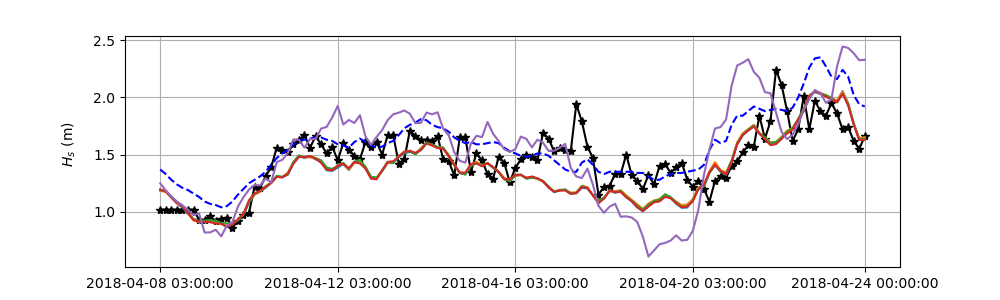}
     \end{subfigure}
        \caption{Comparison between observed data, NOAA numerical simulation and each of the neural network architecture that made up the ensemble of this work. The legend is the same for all three figures. Upper: Itajaí, Middle: Santos, Bottom: Fortaleza.}
        \label{fig:5}
\end{figure}

\begin{figure}[!ht]
     \centering
     \begin{subfigure}{\textwidth}
         \centering
         \includegraphics[width=\textwidth]{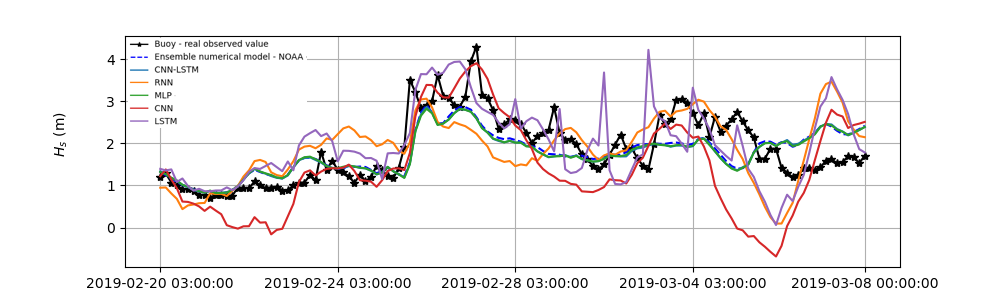}
     \end{subfigure}
     \hfill
     \begin{subfigure}{\textwidth}
         \centering
         \includegraphics[width=\textwidth]{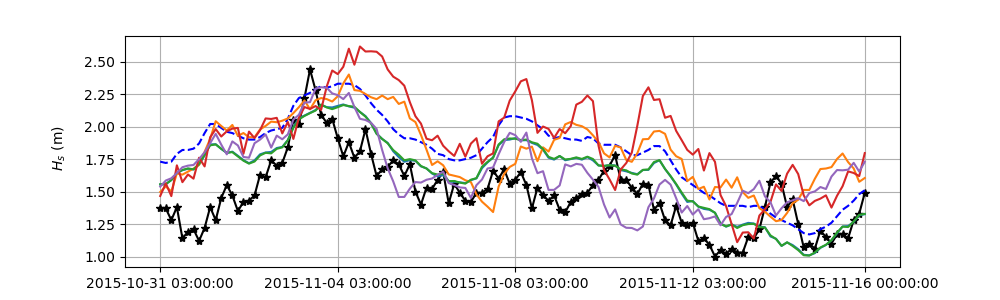}
     \end{subfigure}
     \hfill
     \begin{subfigure}{\textwidth}
         \centering
         \includegraphics[width=\textwidth]{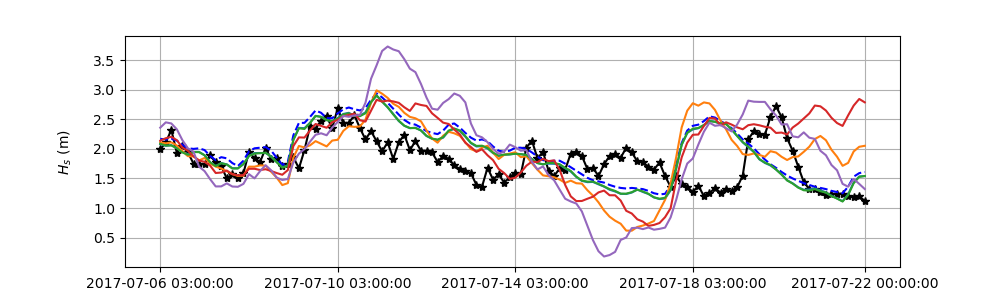}
     \end{subfigure}
        \caption{Comparison between observed data, NOAA numerical simulation and each of the neural network architecture that made up the ensemble of this work. The legend is the same for all three figures. Upper: Rio Grande, Middle: Recife, Bottom: Vitória.}
        \label{fig:6}
\end{figure}

To analyze deeper the results of the ensemble proposed in this work, Figs. \ref{fig:5} and \ref{fig:6} show, at each buoy location, the results separately for the different architectures of neural network, compared with both observational data and NOAA numerical model, and Table \ref{tab:2} display the MAPE metrics. Simulation for buoy location at Itajaí shows that the none of the neural networks were able to capture the two peaks in $H_s$ that occur in the time series, nor does the numerical model. However, for buoy location at Santos, the first peak is well described. These results also show that each architecture of neural networks has its own pattern of learning, since, despite for the buoy results at Itajaí, the results for all other buoys locations have very different behaviours. For instance, for the buoy location Santos, CNN and LSTM  architectures are the ones that follow closely the real data, capturing the peak and valley that occur later in the end of the prediction period, albeit LSTM does not capture well the beginning of the forecast.

\begin{table}
    \centering
    \begin{tabular}{c|cccccc}
       Buoy  &  CNN-LSTM & RNN & MLP & CNN & LSTM &  NN Ensemble \\
       \hline
       Itajaí  & 16.67 & 16.73 & 16.94 & 16.77 & 20.35 &  17.32 \\
       Santos  & 23.55  & 26.65 & 23.53 & 19.48 & 22.44 & 19.23\\
       Fortaleza  & 10.92 & 10.77 & 10.69 & 10.96 & 18.54 & 11.29 \\
       Rio Grande  & 24.14 & 37.60 & 24.34 & 41.43 & 35.66 & 23.80\\
       Recife  & 15.44 & 25.69 & 15.50 & 30.20 & 18.28 &  19.77 \\
       Vitória & 22.24 & 31.05 & 22.26 & 33.79 & 36.87 &  27.72 \\
       \hline
    \end{tabular}
    \caption{MAPE metrics for the different neural networks architecture that made up the ensemble of this work at each buoy location. Values are given in percentage.}
    \label{tab:2}
\end{table}

Now, while CNN shows to be a good alternative based of the locations of Fig. \ref{fig:5}, the same pattern is not shown to locations displayed in Fig. \ref{fig:6}, where this architecture has a lower accuracy. Nevertheless, for these later locations, MLP has one of the best behaviours compared to observational data, also seen in the MAPE metrics. All these differences lead to the conclusion that the neural networks, as well as the ensemble, are location dependent. We can see that for buoy location at Santos and Rio Grande, the ensemble methodology shows real improvement, but for the others, we could choose a specific single network. From an operational perspective for our proposed methodology, depending on the location, some neural network architectures should be removed from the ensemble calculation.

All these differences lead to the conclusion that the ensemble of neural networks are a good alternative to improve the final results.

Unfortunately, due to the characteristic of being black box models, we cannot infer certainly the reasons or the physical meaning of each network result. This is one of the major drawbacks of dealing with classical neural networks algorithms: the lack of physical understanding. Table \ref{tab:2} also does not infer any kind of pattern in the metrics, since the same architecture can yield good accuracy for one location, but a worst in other. The difference here is mostly depending on the data. One of the best results by MAPE accuracy of the ensemble, at Santos, has the particularity of being in a deeper water location than the others, which can explain why at this point our work followed better the observational data. 

\begin{figure}[!ht]
     \centering
     \begin{subfigure}{\textwidth}
         \centering
         \includegraphics[width=\textwidth]{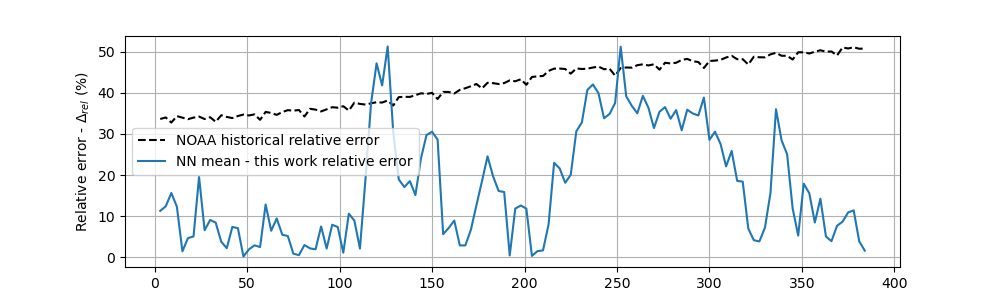}
     \end{subfigure}
     \hfill
     \begin{subfigure}{\textwidth}
         \centering
         \includegraphics[width=\textwidth]{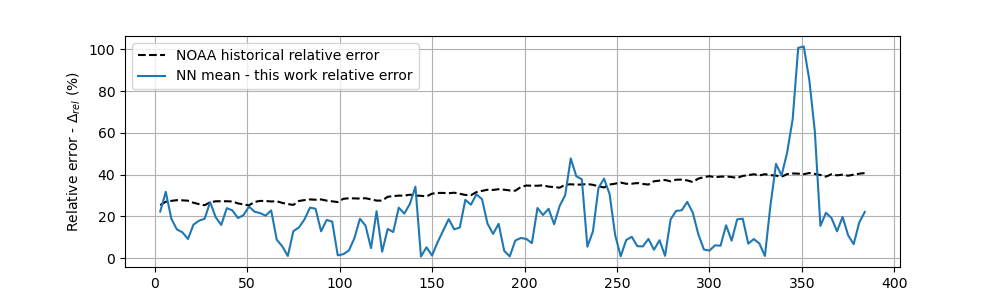}
     \end{subfigure}
     \hfill
     \begin{subfigure}{\textwidth}
         \centering
         \includegraphics[width=\textwidth]{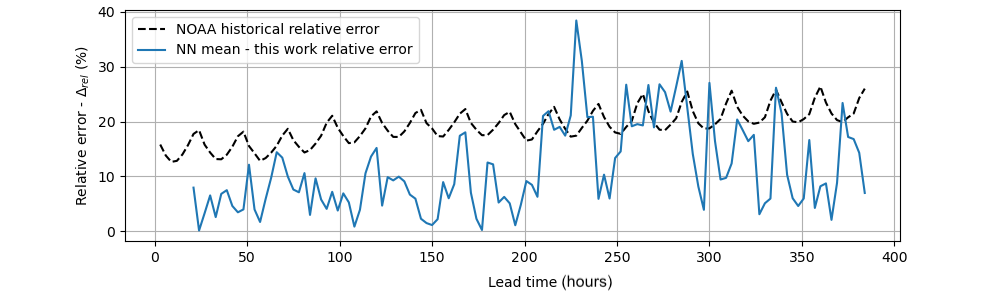}
     \end{subfigure}
        \caption{Comparison between the historical relative error of NOAA numerical model and the ensemble of neural networks, for each lead time from 3 to 384 hours (sixteen days).  Upper: Itajaí, Middle: Santos, Bottom: Fortaleza.}
        \label{fig:7}
\end{figure}

\begin{figure}[!ht]
     \centering
     \begin{subfigure}{\textwidth}
         \centering
         \includegraphics[width=\textwidth]{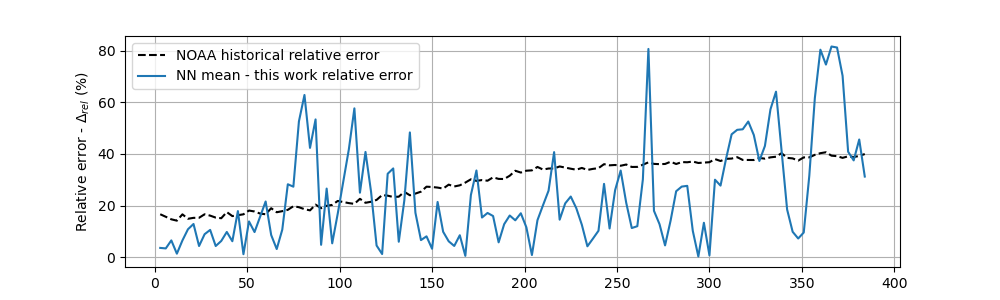}
     \end{subfigure}
     \hfill
     \begin{subfigure}{\textwidth}
         \centering
         \includegraphics[width=\textwidth]{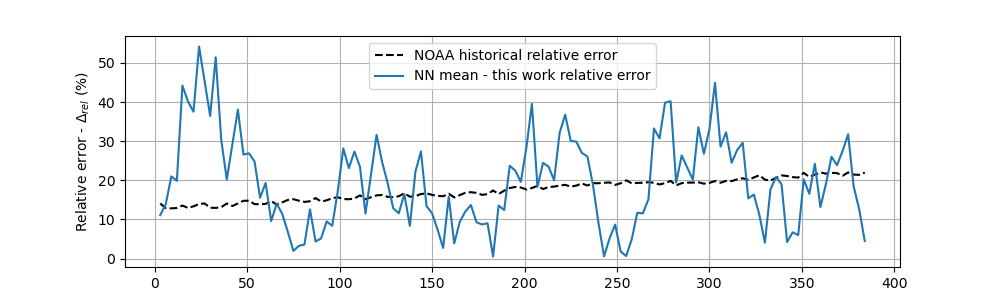}
     \end{subfigure}
     \hfill
     \begin{subfigure}{\textwidth}
         \centering
         \includegraphics[width=\textwidth]{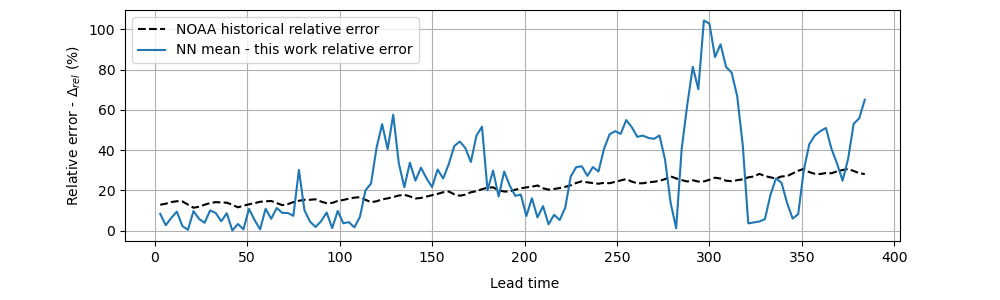}
     \end{subfigure}
        \caption{Comparison between the historical relative error of NOAA numerical model and the ensemble of neural networks, for each lead time from 3 to 384 hours (sixteen days). Upper: Rio Grande, Middle: Recife, Bottom: Vitória.}
        \label{fig:8}
\end{figure}

Since the direct comparison between the ensemble developed in this work and the NOAA numerical model does not reveal significant differences, we analyse the historical behaviour of the numerical forecast. To do this, we calculate the relative error against observational data, i.e., we select the historical data within the same range used for training the neural networks such as the 3-hours lead forecast, average it and assess the error. This process is repeated for each available lead time in the database, ranging from 3 to 384 hours (equivalent to sixteen days). Subsequently, we compare this series of historical mean errors produced by the numerical model with the relative error generated by the ensemble of neural networks, calculated for the specific events studied in this work, as displayed in Figs. \ref{fig:7} and \ref{fig:8}.

The behaviour of NOAA numerical model's historical average error, as can be seen in Figs. \ref{fig:7} and \ref{fig:8} aligns with expectations. As the lead time increases, the error becomes larger. In contrast, the ensemble neural networks exhibits significant variance in the graph because it is not an average such as the historical error. Notably, for buoy locations such as Itajaí, Santos, Fortaleza and Vitória, the relative error of the ensemble developed in this work are essentially lower across the entire domain. Although there is an increase in error in the latter half of the lead time domain (150 to 384), it consistently remains equal to or lower than the historical average error, despite occasional peaks that the ensemble could not describe properly. This shows the neural networks' ability to operate independently of specific time periods or seasonal pattern. They can provide reliable forecasts based solely on historical data and maintain performance throughout the prediction domain.

\begin{figure}[!ht]
     \centering
     \begin{subfigure}{.48\textwidth}
         \centering
         \includegraphics[width=\textwidth]{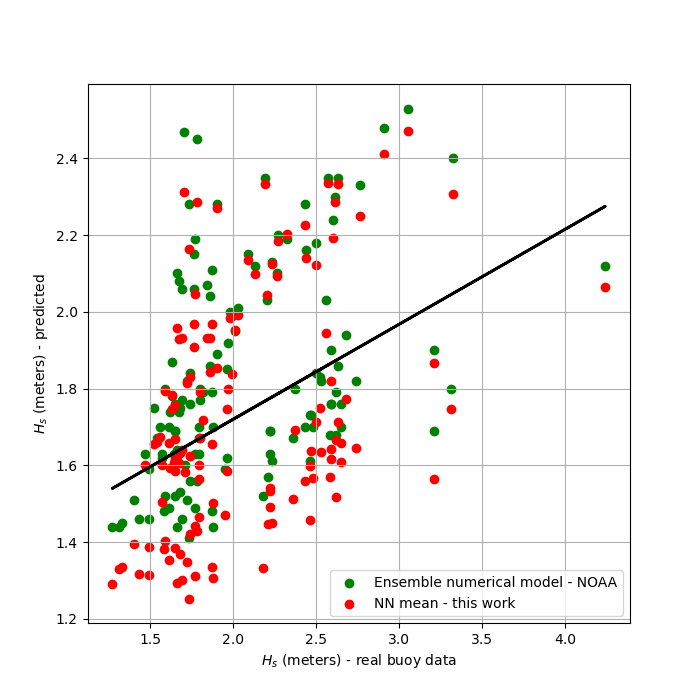}
         \caption{Itajaí}
     \end{subfigure}
     \hfill
     \begin{subfigure}{.48\textwidth}
         \centering
         \includegraphics[width=\textwidth]{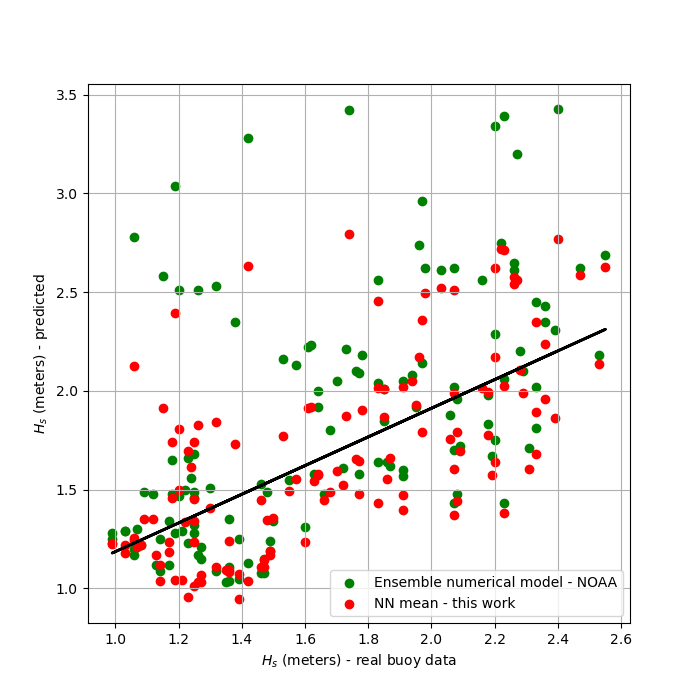}
         \caption{Santos}
     \end{subfigure}
     \hfill
     \begin{subfigure}{.48\textwidth}
         \centering
         \includegraphics[width=\textwidth]{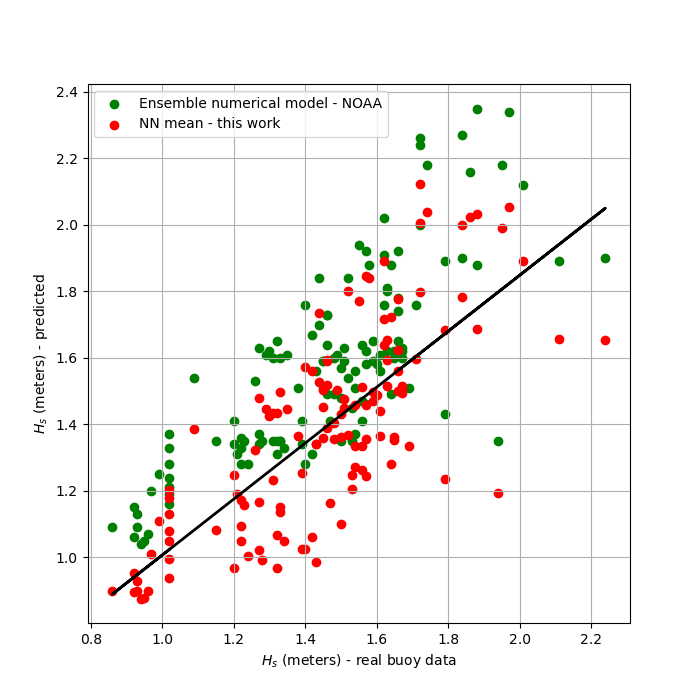}
         \caption{Fortaleza}
     \end{subfigure}
     \hfill
     \begin{subfigure}{.48\textwidth}
         \centering
         \includegraphics[width=\textwidth]{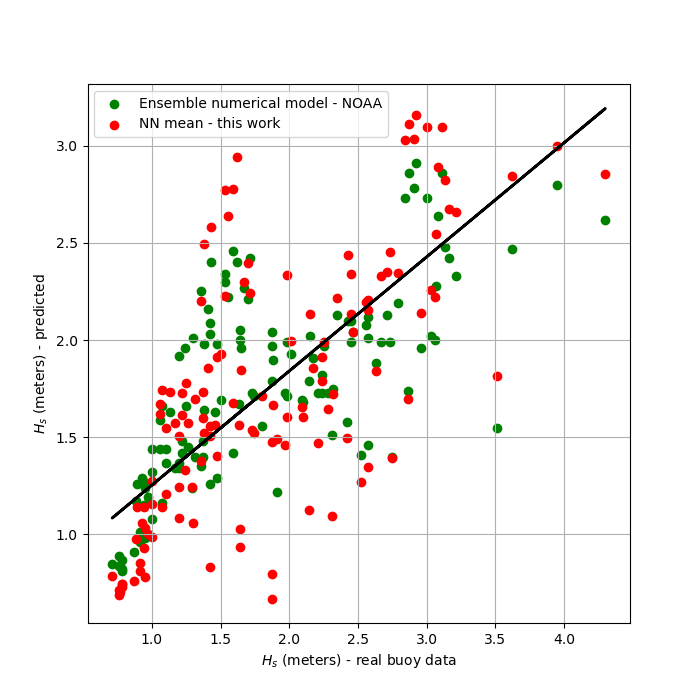}
         \caption{Rio Grande}
     \end{subfigure}
     \hfill
     \begin{subfigure}{.48\textwidth}
         \centering
         \includegraphics[width=\textwidth]{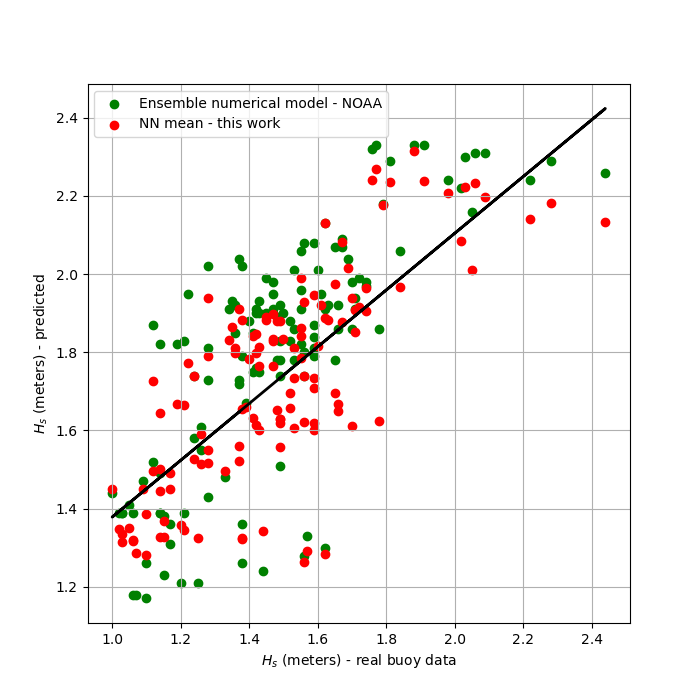}
         \caption{Recife}
     \end{subfigure}
     \hfill
     \begin{subfigure}{.48\textwidth}
         \centering
         \includegraphics[width=\textwidth]{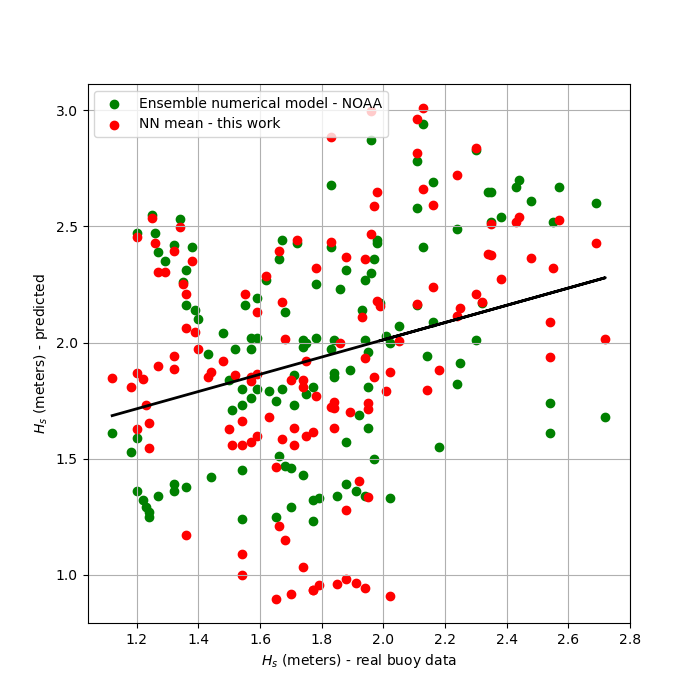}
         \caption{Vitória}
     \end{subfigure}
        \caption{Scatter plots comparing NN ensemble methodology and NOAA ensemble numerical model against real buoy observations. Black line represents the best fit for the data.}
        \label{fig:9}
\end{figure}

\clearpage

Figure \ref{fig:9} shows the scatter plots comparing the NN ensemble methodology developed in this work and the NOAA ensemble numerical forecast, calculated against the real buoy observations. One can also see for each plot, the best fit polynomial of the data. Fortaleza and Recife are the buoys locations where there is a best fit, which follows the model's metrics. The others locations, as can be seen, have a more dispersed pattern of data. The scatter plot for Santos shows that the NN ensemble (red dots) higher values of $H_s$ are closer to the best fit if compared to NOAA numerical forecast, which explains the $5.5\%$ difference in MAPE metric. Besides, the fact that neither this work model nor NOAA forecast could predict the higher peaks for buoy location at Itajaí can be seen in the scatter plot, with higher values away from the best fit polynomial. As expected, consistent with the MAPE metric, buoy location at Vitória brings the most dispersed pattern amidst the results.

There are some technical improvements that could be addressed to increase the neural networks results. For instance, the model would benefit from adding more physical variables that can better represent the sea state. The hyper-parameters of the model, such as number of layers and neurons, can also be optimized to enhance accuracy. These issues will be considered in future works to improve the forecast.

\section{Conclusions} \label{conclusion}

We propose in this work a surrogate methodology to traditional numerical models creating an ensemble of different architectures of artificial neural networks. We base our idea on the well established strategy of using an ensemble of numerical simulations with different initial conditions to produce an accurate result. The neural networks were training with numerical data from NOAA Wave Ensemble Reforecast and target the residual between real observational data and the numerical model output. A new approach for constructing the training and target datasets is also presented.

The results shows that our framework have a good accuracy with metrics that are comparable or, for some cases, superior than the NOAA numerical model. Also, the neural networks ensemble does not reproduce the behaviour of loosing accuracy as the lead time forecast increase, a well known drawback of numerical models. Comparing our result with the historical error of NOAA numerical data for each lead time, we also see an improvement in the performance. The difference in the results between each of the neural network architecture also shows that the strategy of using an ensemble was appropriate. Another major contribution of the present work is that it is the first one to use NOAA Wave Ensemble reforecast data, a large dataset that carries real information on the decay of skill as a function of the forecast lead time, which allows a better discussion about prediction. 

Although our model gives highly accurate predictions, there are some limitations in the results, such as the forecast of peaks. From the neural networks perspective, the architectures that behaved poorly in the simulations should be removed from the set to improve the ensemble results. Since there is not a pattern on which architecture is worst for each location, we want to show in this work that the ensemble methodology can improve, if the right networks are chosen. Besides, as mentioned in the text, we considered numerical simulations from a global wave model, in coastal locations that are not suitable for these.

While surrogate models, such as artificial neural networks, offer notable computational efficiency and straightforward implementation, they do come with a critical limitation – their inherent lack of explainability. This means that we often cannot discern the underlying physical reasons behind the network's decision-making processes when generating results, rendering them as 'black box' models. However, a judicious use of techniques like Physical Informed Neural Networks (PINNs) in frameworks such as the ones employed in this study, opens a promising perspective of research.

\section*{Acknowledgements}
This work has been funded by the Office of Naval Research Global, under the contract No. N629091812124. F. C. Minuzzi also thanks the financial support from the state of Rio Grande do Sul, through FAPERGS, under the term No. 23/2551-0000796-7. We also acknowledge NOAA through Global Ensemble Forecast System (GEFS) and PNBOIA - Brazilian Navy for providing the necessary data.

\bibliographystyle{elsarticle-num} 
\bibliography{mybibfile}

\end{document}